\documentclass[aps,prb,twocolumn,floatfix]{revtex4}
\usepackage{graphicx,bm}
\begin{document}
\pagestyle{plain}
\title{Theory of Current-Induced Magnetization Precession}
\author{M.D. Stiles}
\affiliation{National Institute of Standards and
Technology, Gaithersburg, MD 20899-8412}
\author{Jiang Xiao and A. Zangwill}
\affiliation{School of Physics, Georgia Institute of
Technology, Atlanta, GA 30332-0430}

\begin{abstract} We solve appropriate drift-diffusion and
Landau-Lifshitz-Gilbert equations to demonstrate that unpolarized
current flow from a non-magnet into a ferromagnet can produce a
precession-type instability of the magnetization. The fundamental
origin of the instability is the difference in conductivity
between majority spins and minority spins in the ferromagnet. This
leads to spin accumulation and spin currents that carry angular
momentum across the interface. The component of this angular
momentum perpendicular to the magnetization drives precessional
motion that is opposed by Gilbert damping. Neglecting magnetic
anisotropy and magnetostatics, our approximate analytic and exact
numerical solutions using realistic values for the material
parameters show (for both semi-infinite and thin film geometries)
that a linear instability occurs when both the current density and
the excitation wave vector parallel to the interface are neither
too small nor too large. For many aspects of the problem, the
variation of the magnetization in the direction of the current
flows makes an important contribution.
\end{abstract}

\date{\today}
\maketitle

\section{Introduction}

Recently, the phenomenon of giant magnetoresistance has been used
to unambiguously detect current-induced magnetization reversal in
thin magnetic heterostructures.
\cite{Myers:1999,Grollier:2001,Wegrowe:2001} These observations
confirm theoretical predictions
\cite{Slonczewski:1996,Berger:1996} that angular momentum can be
transferred from current-carrying electrons to the magnetization
of a ferromagnetic film. Earlier reports \cite{others} of the
observation of this so-called {\it spin-transfer} effect
attributed a magnetic origin (magnetization reversal, free
precession, or spin wave excitation) to anomalies seen in
transport measurements. All these experiments used a
ferromagnet/non-magnet/ferromagnet spin valve geometry because a
``polarizer'' ferromagnet was deemed necessary to ensure that the
electrons incident on the ``analyzer'' ferromagnet from the
non-magnet would carry a spin-polarized current. Now, two new
experimental facts have appeared that bear directly on these
issues. First, current injection into a single ferromagnetic layer
has been observed to produce resistance anomalies very similar to
those seen in the spin valves.\cite{Ji:2003} Second, microwave
emission has been detected in at least some spin valve samples
when spin-transfer resistance anomalies occur. \cite{Cornell:2003}

\begin{figure}
\includegraphics{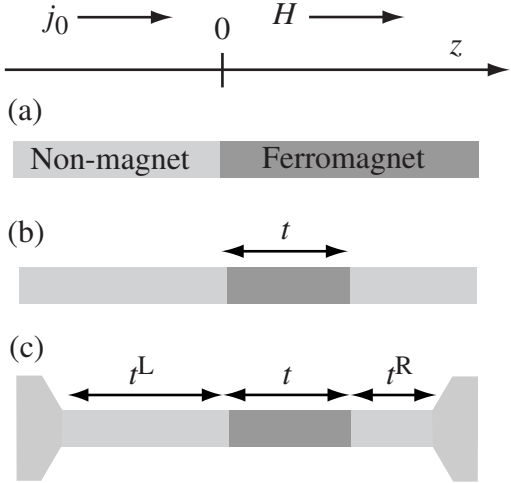}
\caption{The external magnetic field and particle current both
point in the positive $z$-direction for (a) a single-interface
geometry; (b) a thin film geometry; and (c) an electron reservoir
geometry. The interfacial areas are infinite in all cases. In (b),
the non-magnetic leads are semi-infinite in length. In (c), the
non-magnetic leads are attached to electron reservoirs at
equilibrium.} \label{fig:geometry}
\end{figure}

Based on the foregoing, it is natural to inquire whether
magnetic precession occurs
when an unpolarized current flows (perpendicular to
the interface) from a non-magnetic metal into a ferromagnetic metal.
We investigate this question here by performing a linear stability
analysis of a presumptive current-induced precessional state. Our approach to this
problem of coupled transport and magnetization dynamics derives from
previously published work by two of us.
One earlier paper \cite{SZ:2002a} discussed the
quantum mechanical origin of
spin-transfer torque at a ferromagnetic/non-magnetic interface. Another
paper \cite{SZ:2002b} solved a matrix Boltzmann equation to compute spin currents, spin
accumulation, magnetoresistance, and spin-transfer torques
for multilayers.  In the present work, we study both single and multiple interface geometries
in the presence of an external magnetic field (normal to the interface) strong enough
to saturate the magnetization (Figure~\ref{fig:geometry}). Magnetic anisotropy and
magnetostatics are neglected for simplicity.
Our approximate analytic and exact
numerical solutions reveal that an initial instability toward precession
occurs when both the current
density and the  excitation
wave vector parallel to the interface are neither too large nor too small.
Otherwise, Gilbert damping suppresses the instability. An important
feature of our model is that the
magnetization is permitted to vary in the direction normal to the
interface. Among other things, this means that precession can occur in the film geometry even when
the two interfaces are identical.
Other systematic
features of the precessional state will be described in detail below.

Our results complement and extend previous
theoretical studies of current-induced precession in magnetic films.
In their pioneering work, Slonczewski \cite{Slonczewski:1996} and Berger \cite{Berger:1996} used
analytic model calculations to predict the critical
current for the onset of precession in spin valve structures with uniform but
non-collinear magnetizations in the two ferromagnetic layers.
Recent work shows that the seemingly different answers they obtained
are not inconsistent. \cite{Tserkovnyak:2003}
Two global studies of precession and switching
in single-domain spin valves \cite{Bazaliy:2001} and single-domain particles \cite{Sun:2000}
supplement the Landau-Lifshitz-Gilbert equation with the spin-transfer
torque derived in Ref.~\onlinecite{Slonczewski:1996}.
Bazaliy et al. \cite{Bazaliy:1998} relaxed the assumption of uniform magnetization
and studied the effect of a legislated spin current flowing from a non-magnet into
a semi-infinite ferromagnet. They found that
spontaneous excitation of spin waves occurs when the current density is
sufficiently large. Finally, we draw attention to a very recent application
\cite{Polianski:2003} of magnetoelectronic circuit theory \cite{Brataas:2001}
to study current-induced precession in an
ultrathin ferromagnetic film with uniform magnetization. For unpolarized current
densities above a critical value, the authors
predict that unstable spin wave modes exist only for non-zero
wave vectors parallel to the interface. Our results
confirm this prediction.

The calculations presented in this paper combine phenomenological constitutive equations for the
electric current and the spin current (drift-diffusion equations) with
a Landau-Lifshitz-Gilbert equation generalized to include spin-transfer torque. The boundary
conditions at
each non-magnet/ferromagnet interface are chosen so that relevant observables reproduce
the results of a Boltzmann equation treatment of the same problem. Using experimental data
to fix the parameters, we
choose a value for the electric current density, and solve for the interface charge density
(chemical potential discontinuity),
spin accumulation, and spin current assuming
a precessing magnetization with an amplitude and phase that depends on the
running variable $z$ in Figure~\ref{fig:geometry}. The
presumed precession is a legitimate mode of the system if the computed spin-transfer torque
balances the ferromagnetic exchange torque at each interface. \cite{Bazaliy:1998}
The mode is unstable (spontaneous excitation in the presence of a current) if
the imaginary part of the mode frequency is negative. The mode is stable and uninteresting
if Gilbert damping overwhelms the effect of spin transfer.

The plan of this paper is as follows. In Section II, we set down
the basic definitions and equations of the theory. Section III
gives a qualitative picture of the origin of the precessional
instability. The boundary conditions for the transport and
magnetization equations are given in Section IV. Sections V and VI
present the details of the calculation for a single interface
geometry and a thin film geometry, respectively. Section VII
discusses our results. Section VIII examines the effect of
non-magnetic leads using a reservoir
geometry.\cite{Polianski:2003} Section IX discusses some
limitations of our model and suggestions for future work. Section
X is a summary. Appendix A derives the effective exchange field
for a bounded ferromagnet. The transport boundary equations are
derived in Appendix B.

\section{Definitions \& Basic Equations}

In an earlier paper, \cite{SZ:2002a} we defined the fundamental variables of spin transport
quantum mechanically and
described the various relations that exist among them. Here, we proceed phenomenologically
and decompose the total
number density of conduction electrons into majority and minority contributions
$n=n_{\uparrow}+n_{\downarrow}$ so that the current-induced deviations of these quantities from their
equilibrium values are related by
\begin{eqnarray}
\delta n = \delta n_{\uparrow} + \delta n_{\downarrow}.
\label{deltan}
\end{eqnarray}
In the drift-diffusion approximation, the particle
current in the {\it non-magnet} is
\begin{eqnarray}
{\bf j}^{\rm NM}= (\sigma/e) {\bf E} - D \bm{\nabla} \delta n,
\label{Fick}
\end{eqnarray}
where ${\bf E}$ is the electric field, $\sigma$ is the
conductivity, and $D$ is the electron diffusion constant. We speak
of {\it spin accumulation} in the non-magnet when the spin density
\begin{eqnarray}
{\bf m} = {\hbar \over 2}(\delta n_{\uparrow} - \delta n_{\downarrow})\hat{\bf m}
\label{spindensity}
\end{eqnarray}
is different from zero.  Notice that
we use the particle current and the spin density as variables rather than
the charge current and the magnetization.

A so-called {\it spin current density}
\begin{equation}
Q^{\rm NM}_{\alpha \beta}=-D\nabla_{\beta}m_{\alpha}
\label{spincurrent}
\end{equation}
flows in the non-magnet when the spin accumulation is not uniform
in space. In (\ref{spincurrent}), the first index of
$Q_{\alpha,\beta}^{\rm NM}$ labels the $\alpha^{\rm th}$ Cartesian
component of the electron spin vector while the second index
labels the $\beta^{\rm th}$ Cartesian component of the flow
direction.

In the {\it ferromagnet}, we make explicit use of the fact that
there is no appreciable spin accumulation or spin current in the
direction perpendicular to the magnetization.
\cite{SZ:2002a,Shpiro:2003} Accordingly, $\delta m = (\hbar
/2)(\delta n_{\uparrow}-\delta n_{\downarrow})$ is the spin
accumulation so
\begin{equation}
{\bf M}=(m_s+\delta m){\bf u}
\label{ferromagnetization}
\end{equation}
is the total spin density. Here, ${\bf u}$ is a unit vector,
$m_{\rm s}=\hbar M_{\rm s}/g \mu_B$, and $M_{\rm s}$ is
the saturation magnetization.
We adopt a two-current model \cite{Fert:1969} so ${\bf j}^{\rm FM}={\bf j}_{\uparrow}+{\bf j}_{\downarrow}$
where
\begin{eqnarray}
{\bf j}_{\uparrow} &=&(\sigma_{\uparrow}/e){\bf E}-D_{\uparrow}{\bm \nabla}
\delta n_{\uparrow} \nonumber \\ \nonumber \\
{\bf j}_{\downarrow} &=&(\sigma_{\downarrow}/e){\bf E}-D_{\downarrow}{\bm \nabla}
\delta n_{\downarrow},
\label{updowncurrent}
\end{eqnarray}
using an obvious notation for the conductivity and diffusion constants of majority and
minority spin electrons. Combining all the above, the total particle current in the ferromagnet is
\begin{equation}
{\bf j}^{\rm FM}={\sigma_{\uparrow}+\sigma_{\downarrow}\over e}{\bf E}
-{D_{\uparrow}+D_{\downarrow}\over 2}{\bm \nabla} \delta n
-{D_{\uparrow}-D_{\downarrow}\over \hbar}{\bm \nabla} \delta m.
\label{ferrocurrent}
\end{equation}

The corresponding spin current density in the ferromagnet is
\begin{equation}
Q^{\rm FM}_{\alpha \beta}=\textstyle{1 \over 2}\hbar u_{\alpha}
({\bf j}_{\uparrow}-{\bf j}_{\downarrow})_\beta.
\label{Qdef}
\end{equation}
Consistent with our neglect of transverse spin accumulation in (\ref{ferromagnetization}), this
formula presumes that the spin part of the spin
current in a ferromagnet is always parallel to the spin density.
Then,
using (\ref{updowncurrent}), we get
\begin{eqnarray}
Q_{\alpha \beta}^{\rm FM} &=& {\hbar \over 2}{\sigma_{\uparrow}-\sigma_{\downarrow}\over e}u_{\alpha}E_{\beta}
-{\hbar \over 2}{D_{\uparrow}-D_{\downarrow}\over 2}u_{\alpha}\nabla_{\beta} \delta n \nonumber \\
& & \mbox{} - {D_{\uparrow}+D_{\downarrow}\over 2}u_{\alpha}\nabla_{\beta} \delta m.
\label{ferrospincurrent}
\end{eqnarray}

Textbooks show that the charge density inside a current-carrying
Ohmic conductor is zero (except within a screening distance of an
interface or free surface). Therefore, we are justified to put
$\delta n=0$  in (\ref{Fick}), (\ref{ferrocurrent}), and
(\ref{ferrospincurrent}). We will also fix the current to be ${\bf
j}=j_0\hat{z}$ everywhere. Using this fact in
(\ref{ferrocurrent}), we can eliminate ${\bf E}$ from
(\ref{ferrospincurrent}) and simplify the components of the
ferromagnetic spin current density that flow in directions
longitudinal ($z$) and transverse ($\perp$) to the interface
normal. The result is
\begin{eqnarray}
Q_{\alpha z}^{\rm FM} &=&
\textstyle{1 \over 2}\hbar\gamma^{\rm p} u_\alpha
 j_0
-D_z   u_\alpha \nabla_z \delta m
\nonumber\\ \nonumber \\
Q_{\alpha \perp}^{\rm FM} &=&
-D_{\perp}  u_\alpha \nabla_\perp \delta m,
\label{Qinmagnet}
\end{eqnarray}
where $\gamma^{\rm p}=(\sigma_\uparrow -\sigma_\downarrow)/
(\sigma_\uparrow + \sigma_\downarrow)$
is the degree of current polarization desired by the ferromagnetic bulk,
$D_z=(D_\uparrow \sigma_\downarrow + D_\downarrow \sigma_\uparrow)/
(\sigma_{\uparrow}+\sigma_{\downarrow})$ is an effective longitudinal
spin diffusion constant, and
$D_\perp=\textstyle{1\over 2}(D_\uparrow + D_\downarrow)$ is an
effective transverse spin diffusion constant.

The equation of motion for the spin density in the {\it non-magnet}
is a continuity equation for the spin accumulation. \cite{SZ:2002a} Taking account of
distributed spin-transfer torque,
spin flip scattering and torque
from the external magnetic field, we find
\begin{equation}
\dot{m}_{\alpha}+\nabla_{\beta}Q^{\rm NM}_{\alpha \beta} +\gamma \left[{\bf m} \times
{\bf H}_{\rm ext}\right]_{\alpha}+{m_{\alpha} \over \tau_{\rm sf}}=0,
\label{LLGnm}
\end{equation}
where $\tau_{\rm sf}$ is the spin-flip scattering time in the non-magnet and
$\gamma {\bf H}=(g \mu_B / \hbar){\bf B}$.
The equation of motion for the spin density in the {\it ferromagnet} is
the Landau-Lifshitz-Gilbert equation supplemented by
a term to account for spin-transfer torque as in (\ref{LLGnm}). Specifically,
\begin{eqnarray}
\dot{M}_{\alpha}
+ \nabla_{\beta}Q_{\alpha \beta}^{\rm FM} +\gamma \left[{\bf M}\times{\bf H}_{\rm eff}\right]_{\alpha}
+{u_{\alpha}\delta m \over \bar{\tau}_{\rm sf}}=0,
\label{LLGm}
\end{eqnarray}
where $\bar{\tau}_{\rm sf}$ is the spin-flip scattering time in the ferromagnet.
The effective field ${\bf H}_{\rm eff}$ accounts for the external field, exchange, and damping:
\begin{eqnarray}
{\bf H}_{\rm eff} &=& {\bf H}_{\rm ext} + {\omega_0 l^2_{\rm ex}  \over \gamma m_{\rm s}}
\nabla^2 {\bf M}
-  {\alpha \over \gamma |{\bf M}|} \dot{\bf M}.
\label{Heff}
\end{eqnarray}
Here,  $l_{\rm ex}=\sqrt{2A_{\rm ex}/\mu_0M^2_s}$ is the exchange
length in a ferromagnet with spin stiffness $A_{\rm ex}$,
$\omega_0=\gamma M_{\rm s}$, and $\alpha$ is a phenomenological
damping parameter. We ignore magnetostatics, but for fields and
magnetization perpendicular to the interface, its main effect
is to shift the applied magnetic field by an amount $M_{\rm s}$.

\begin{figure}
\includegraphics[height=8.0in]{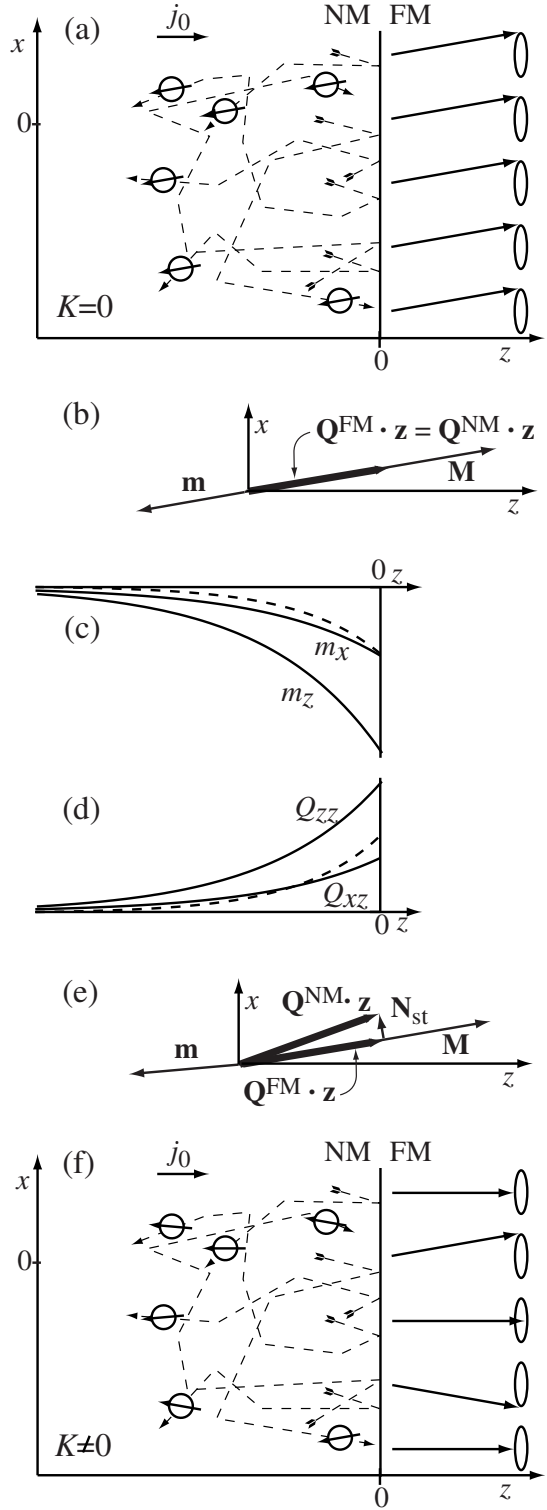}
\caption{(a) Cartoon of steady-state spin configuration when
$K=0$; (b) spin accumulation and spin current when $K=0$;
(c) solid curves: $m_z(x=0,z)$ and $m_x(x=0,z)$ when $K=0$; dashed curve:
$m_x(x=0,z)$ when $K\neq 0$; (d) solid curves: $Q^{\rm NM}_{zz}(x=0,z)$ and
$Q^{\rm NM}_{xz}(x=0,z)$ when $K=0$; dashed curve:
$Q^{\rm NM}_{xz}(x=0,z)$ when $K\neq 0$; (e) same as (b) when $K\neq 0$;
(f) same as (a) when $K\neq 0$. See text for details.}
\label{fig:cartoon}
\end{figure}

\section{Origin of the Instability}
\label{sec:origin}
 Two characteristic features of the spin current
$Q_{\alpha \beta}$ underlie the qualitative physics of
current-induced precession. First, the  structure of the
spin-transfer torque in (\ref{LLGnm}) and (\ref{LLGm}) implies
that the magnetization at the $z=0$ interface in
Figure~\ref{fig:geometry} feels a torque per unit area. It is
\begin{eqnarray}
{{\bf N}_{\rm st}\over A}=
- \lim_{\epsilon \rightarrow 0}\int\limits_{-\epsilon}^{\epsilon} dz
{\bm \nabla}\cdot {\bf Q}
&=&({\bf Q}^{\rm NM}-  {\bf Q}^{\rm FM})\cdot \hat{\bf z} \nonumber \\
&\simeq& {\bf Q}^{\rm NM}_{\perp' z},
\label{Nst}
\end{eqnarray}
where $\perp'$ refers to the directions transverse to the
ferromagnetic magnetization ($z'$ is the direction along the
ferromagnetic magnetization).
The last line in (\ref{Nst}) is true because
\begin{equation}
Q^{\rm NM}_{z'z}=Q^{\rm FM}_{z'z}
\label{nQcont}
\end{equation}
and $Q^{\rm FM}_{\perp' z}\simeq0$ at the interface between a non-magnet and
a ferromagnet due to spin-filtering and other effects.
 \cite{Slonczewski:1996,Berger:1996,Waintal:2000,SZ:2002a}

The second bit of information we need is a linear kinetic equation that relates the
transverse
spin current at the interface, ${\bf Q}^{\rm NM}_{\perp' z}$, to the interfacial densities of our problem
(charge accumulation,  spin accumulation in the non-magnet, and
ferromagnetic spin accumulation). Of these, for a static magnetization
only the spin accumulation ${\bf m}$ in (\ref{spindensity}) has a transverse component. Therefore,
it must be the case that
\begin{equation}
{\bf Q}^{\rm NM}_{\perp' z} = b \,{\bf m}_{\perp '},
\label{nQBC}
\end{equation}
where $b$ is a constant. More detailed considerations (see Appendix B) show that $b>0$.

For simplicity, we will study (in this section) a {\it
time-independent} situation where electrons flow from a non-magnet
into a ferromagnet with a small amplitude, frozen spin wave at the
interface. That is, in (\ref{ferromagnetization}), we let ${\bf
u}=\hat{\bf z}+{\bf u}_\perp \exp(i{\bf K}\cdot {\bf R})$ where
${\bf R}$ and ${\bf K}$ are two-dimensional vectors in the plane
of the interface. This means that conduction electrons incident on
the ferromagnet from the non-magnet see a static magnetization
\begin{equation}
{\bf M}=M\left\{{\rm Re}[ u_\perp e^{i{\bf K}\cdot{\bf R}} ],
{\rm Im}[ u_\perp e^{i{\bf K}\cdot{\bf R}} ],1\right\}.
\label{mf}
\end{equation}

In the non-magnet itself, (\ref{spincurrent}) and (\ref{LLGnm}) with ${\bf H}_{\rm ext}=0$ combine
to give
\begin{equation}
D \nabla^2 m_{\alpha} = {m_\alpha \over \tau_{\rm sf} }
\label{NMdiff}
\end{equation}
for the steady-state spin accumulation in the non-magnet.
This is solved by
\begin{equation}
{\bf m} =
\left\{ {\rm Re}[ m_\perp (z) e^{i{\bf K}\cdot{\bf R}}] ,
       {\rm Im}[ m_\perp (z) e^{i{\bf K}\cdot{\bf R}}] ,
       m_z(z) \right\},
\label{ansatz}
\end{equation}
if
\begin{eqnarray}
m_z(z) & =& m_z \exp(z/l_{\rm sf}) \nonumber \\
m_\perp(z)& =& m_\perp \exp(\kappa z)
\label{nmzz}
\end{eqnarray}
with $l_{\rm sf}^{2}= D \tau_{\rm sf}$  and $l_{\rm sf}^2 \kappa^2
= 1+l_{\rm sf}^2 K^2.$
\\

We begin our analysis with the right side of the top panel in
Figure~\ref{fig:cartoon}. This shows a spin pattern in the
ferromagnet that is spatially uniform ($K=0$). In panel (b), just
below, the arrow labeled ${\bf m}$ reiterates the information
conveyed by the diffusing spins on the left side of panel (a).
Namely, the spin accumulation in the non-magnet is everywhere
anti-parallel to ${\bf M}$. To understand this, we note first that
$Q^{\rm FM}_{zz}>0$ on account of the first term in
(\ref{Qinmagnet}).\cite{Qz'z>0} Therefore, to satisfy
(\ref{nQcont}), there must be a collection of spins in the
non-magnet (polarized parallel to ${\bf M}$) flowing from left to
right toward the interface. We interpret this, equivalently and
consistently with the diffusive solutions (\ref{nmzz})  graphed as
solid curves in panel (c), as a flow of {\it anti-parallel} spins
{\it away} from the interface into the non-magnet. Qualitatively,
we imagine that spins diffuse into the non-magnet after being
``imprinted'' in the anti-parallel direction by the ferromagnet.
Randomly walking spins (including those that happen to diffuse
back to the interface) only encounter spins parallel to the
themselves because all imprinting was done by ferromagnetic spins
that are parallel. From (\ref{Nst}) and (\ref{nQBC}), ${\bf
N}_{\rm st} = A{\bf Q}^{\rm NM}_{\perp ' z}=0$ because ${\bf
m}_{\perp '}=0$. There is no driving force for precession when the
ferromagnetic magnetization is uniform. On the other hand, the
spin currents $Q^{\rm NM}_{zz}$ and ${\bf Q}^{\rm NM}_{\perp z}$
are both non-zero [solid curves in panel (d)] because ${\bf M}$ is
not normal to the interface.

The situation changes markedly when the magnetization ${\bf M}$
varies along the interface ($K\neq 0$) [right side of
Figure~\ref{fig:cartoon}(f)]. As before, spins diffuse into the
non-magnet after being imprinted in the direction anti-parallel to
the local ferromagnetic magnetization. However, because the
imprinting ferromagnetic spins now differ in their transverse
directions, spins diffusing in the non-magnet typically encounter
spins whose transverse component differs from their own. The net
effect is a reduction in the transverse spin accumulation [dashed
curve in panel(c)]. The effect is greater farther away from the
interface because the transverse diffusion is superimposed on
longitudinal diffusion away from the interface.  From
(\ref{spincurrent}), there is a corresponding increase in the
transverse spin current near the interface [dashed curve in panel
(d)]. Compared to panel (b), the spin accumulation vector ${\bf
m}$ in panel (e) bends closer the negative $z$-axis and thus
acquires a positive component perpendicular to ${\bf M}$.
Therefore, from (\ref{Nst}) and (\ref{nQBC}), the vectors ${\bf
Q}^{\rm NM}$ and ${\bf N}_{\rm st}$ appear as sketched in panel
(e). The indicated torque tends to rotate  ${\bf M}$ away from the
interface normal. That is, ${\bf N}_{\rm st}$ tends to {\it
increase} the amplitude ${\bf u}$ of the incipient spin wave
assumed at the beginning of the discussion. This is the
fundamental origin of the precessional instability. Thus, in
contrast to the common spin-valve geometry, we have here a single
ferromagnet that both generates a spin current and is acted on by
that spin current.

\begin{figure}
\includegraphics{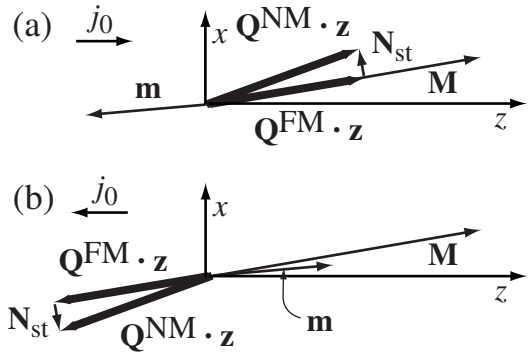}
\caption{(a) Interfacial spin current and spin for current flow
from the non-magnet into the ferromagnet. This diagram is the same
as Figure~\ref{fig:cartoon}(e); (b) Same as (a) except that the
current flows from the ferromagnet into the non-magnet.}
\label{fig:reversecurrent}
\end{figure}

Two additional qualitative points emerge form the discussion
above. First, if the current $j_0$ reverses direction, so
electrons flow from the ferromagnet into the non-magnet,
(\ref{Qinmagnet}) tells us that the sign of $Q_{zz}$
reverses.\cite{Qz'z>0} The direction of the spin accumulation
reverses also so (\ref{spincurrent}) remains correct. The decrease
in transverse spin accumulation which rotates ${\bf m}$ toward the
$z$-axis now implies that the direction of ${\bf N}_{\rm st}$
reverses as well (Figure~\ref{fig:reversecurrent}). This torque
tends to reduce the amplitude of the presumed spin wave. We
conclude that electron flow from a ferromagnet into a non-magnet
tends to {\it suppress} a precessional instability. This type of
non-reciprocal behavior with respect to the direction of current
flow is a characteristic feature of spin-transfer phenomena.
\cite{Slonczewski:1996}

Finally, we remark that the presence of a second ``polarizer''
ferromagnetic film upstream in Figure~\ref{fig:cartoon} tends to
enhance the precessional instability. Even the $K=0$ mode can
become unstable as long as the magnetization of the polarizer is
not collinear with the magnetization of the ``analyzer''. This is
so because, in that case, the electrons incident on the $z=0$
plane  carry a spin current with a non-zero component in a
direction perpendicular to ${\bf M}$. As a practical matter, this
means that the critical current for the onset of precession is
lower when two ferromagnets are present than for the case of a
single ferromagnet studied by us.

\section{Boundary Conditions}

A quantitative solution of our problem requires boundary
conditions for the transport and magnetization equations set down
in Section II. One of these arises because the exchange
interaction is effective right up to the surface of a bounded
ferromagnet. For the thin film geometries of
Figure~\ref{fig:geometry},  we show in Appendix A that an exchange
torque density ${\bf n}_{\rm ex}=-\gamma {\bf M} \times {\bf
H}^{\rm S}$ exists where
\begin{equation}
{\bf H}_{\rm ex}^{\rm S}={\omega_0 l^2_{\rm ex}  \over \gamma m_{\rm s}}
\left[\delta(z)-\delta(z-t)\right]\partial_z{\bf M}.
\label{Hsex}
\end{equation}
This shows that  a localized torque acts on each interface in Figure~\ref{fig:geometry}(b).
For the single interface geometry, the $\delta(z-t)$ term is absent.
Spin wave calculations for bounded ferromagnets \cite{Gurevich:1996}
eliminate the surface torque produced by (\ref{Hsex})
by imposing the boundary condition $\partial_z{\bf M}=0$
at each interface. For our problem, the spin-transfer torque
(\ref{Nst}) also acts at each interface. This dictates that we determine the spin wave modes
by imposing a zero {\it net} torque requirement at each interface.\cite{Bazaliy:1998}
The specific boundary condition is
\begin{equation}
{\bf N}_{\rm ex}+{\bf N}_{\rm st}=0.
\label{zerotorque}
\end{equation}
where ${\bf N}_{\rm ex}$ is the exchange torque density ${\bf
n}_{\rm ex}$ integrated over an infinitesimally thin pillbox as in
(\ref{Nst}).

A distribution of charge/spin current
where the spins precess at a complex
frequency $\omega=\omega_{\rm R} + i \omega_{\rm I}$ is a mode of the system if
(\ref{LLGnm}) is satisfied in non-magnetic regions, (\ref{LLGm}) is satisfied in
ferromagnetic regions, and (\ref{zerotorque}) is satisfied at every interface.
When $\omega_{\rm I}>0$, the mode is stable and uninteresting because its amplitude
never grows large enough to be observed. True current-induced precession corresponds
to modes with $\omega_{\rm I}<0$.

Of course, the currents must also be consistent with the
spin-dependent scattering that occurs at the interface between a
magnet and a non-magnet. Previously, two of the present authors
\cite{SZ:2002b} addressed this issue by imposing scattering
boundary conditions on a $2 \times 2$ matrix Boltzmann equation
and computing all observables from the semi-classical distribution
functions $g_{\uparrow}({\bf k},{\bf r})$ and $g_{\downarrow}({\bf
k},{\bf r})$. In this paper, we apply boundary conditions to the
observables directly in a manner that is consistent with this
Boltzmann equation treatment.

We show in Appendix B that the boundary conditions obtained in this way take the form of a set
of linear kinetic equations which relate the interfacial currents to the interfacial
densities. Here, we quote a form that is valid for small amplitude
precessional motion around ${\bf H}_{\rm ext}=H\hat{z}$. That is, we
assume (here and in all subsequent work) that ${\bf u}=\hat{\bf z}+{\bf u}_{\perp}$
with $u_\perp \ll 1$:
\begin{eqnarray}
j_{0} &=&  \delta n_{0} [ w_0 ( X_\uparrow + X_\downarrow ) ]
       + 2\hbar^{-1}m_{z}[ w_0 ( X_\uparrow - X_\downarrow ) ] \nonumber \\
   & & \mbox{} - 2\hbar^{-1}\delta m
         [ X_\uparrow w_\uparrow - X_\downarrow w_\downarrow ]
\nonumber\\ \nonumber \\
Q^{\rm NM}_{xz} &=& u_x Q^{\rm NM}_{zz} +
w_0 ( m_{x} - u_x m_{z})   \nonumber\\ \nonumber \\
Q^{\rm NM}_{yz} &=&  u_y Q^{\rm NM}_{zz} +
w_0 ( m_{y}  - u_y m_{z} ) \nonumber\\ \nonumber \\
Q^{\rm NM}_{zz} &=&  \textstyle{1\over 2}\hbar\delta n_{0}[ w_0 ( X_\uparrow - X_\downarrow ) ]
       + m_{z} [ w_0 ( X_\uparrow + X_\downarrow ) ] \nonumber \\
& & \mbox{}       - \delta m
         [ X_\uparrow w_\uparrow + X_\downarrow w_\downarrow ].
\label{BC}
\end{eqnarray}
The constant
\begin{equation}
w_0={A_{\rm FS}\over (2\pi)^3\hbar}\left[{1\over 2}{\partial n \over \partial \mu}\right]^{-1},
\label{wdef}
\end{equation}
where $A_{\rm FS}$ is the area of the Fermi surface projected onto the interface. The constants
$w_\uparrow$ and $w_\downarrow$ are defined similarly with the densities of states per spin
$\partial n_\uparrow /\partial \mu_\uparrow$ and $\partial n_\downarrow / \partial \mu_\downarrow$.
Finally,
\begin{eqnarray}
X_\sigma &=& { 1 - {\rm R}_\sigma \over 4 {\rm R}_\sigma },
\label{Xdef}
\end{eqnarray}
where ${\rm R}_{\uparrow}=|R_{\uparrow}|^2$ and ${\rm
R}_{\downarrow}=|R_{\downarrow}|^2$ are the squares of the
majority and minority electron reflection amplitudes averaged over
the Fermi surface. The appearance of the interfacial charge
accumulation $\delta n _0$  on the right side of (\ref{BC}) is not
inconsistent with our earlier statement that $\delta n^{\rm NM} =
\delta n^{\rm FM} = 0$ in the bulk. There is a resistance
associated with the interface whenever ${\rm R}_{\sigma} \neq 0$.
The corresponding voltage drop may be thought of as arising from a
dipole layer of charge difference $\delta n_0$ localized at the
interface.

We note in closing that if ${\bf u}_\perp$ were zero (${\bf M} \parallel \hat{\bf z}$), the second
and third boundary conditions in (\ref{BC}) reduce to (\ref{nQBC}).
Appendix B demonstrates that
the terms proportional to ${\bf u}_\perp$ in (\ref{BC}) account for the fact that ${\bf M}$ is slightly tilted
away from the interface normal.

\section{Single Interface Geometry}
We showed in Section III that spin-transfer torque can induce a
precessional instability. To discover if it actually does so
requires a full solution of the coupled transport/magnetization
dynamics equations with their boundary conditions. In particular,
we need convenient formulae for the interfacial torques that enter
(\ref{zerotorque}). For the single interface  geometry of
Figure~\ref{fig:geometry}(a), a suitable precessional ansatz for
the magnetization in the non-magnet is
\begin{equation}
{\bf m} =
\left\{ {\rm Re}[ m_\perp (z) \Phi] ,
       {\rm Im}[ m_\perp (z) \Phi] ,
       m_z(z) \right\}
\label{mNM}
\end{equation}
where $\Phi=e^{i({\bf K}\cdot{\bf R}+\omega t)}$.
The corresponding ansatz in the ferromagnet is
\begin{equation}
{\bf u}=\hat{\bf z}+{\bf u}_\perp (z) \Phi,
\label{uFM}
\end{equation}
so from ({\ref{ferromagnetization}),
\begin{eqnarray}
{\bf M}
&=& M(z)
\left\{{\rm Re}\left[u_\perp(z)\Phi \right],
{\rm Im}\left[u_\perp(z) \Phi \right],
1 \right\}
\label{mFM}
\end{eqnarray}
where
\begin{equation}
M(z)=m_{\rm s} +\delta m(z).
\label{Mofz}
\end{equation}

\subsection{Interfacial Spin-Transfer Torque}

To find the spin-transfer torque,
we substitute (\ref{mNM}) and (\ref{mFM})
into the longitudinal
component of the equations of motion (\ref{LLGnm})
and (\ref{LLGm}). The solutions are [cf. (\ref{nmzz})]
\begin{equation}
m_z(z)  = m_z \exp (z/l_{\rm sf})
\label{nmm}
\end{equation}
and
\begin{equation}
\delta m(z)=\delta m \exp( -z/\bar{l}_{\rm sf})
\label{mm}
\end{equation}
with $\bar{l}_{\rm sf}^2=D_z \bar{\tau}_{\rm sf}$. These are
used in the
longitudinal component of
(\ref{spincurrent}) and (\ref{Qinmagnet})
to get the interfacial relations
\begin{equation}
 Q^{\rm NM}_{zz}=-(D/l_{\rm sf})m_z= Q_{zz}
\label{QQNM}
\end{equation}
and
\begin{equation}
Q^{\rm FM}_{zz}=
\textstyle{1\over2}\hbar\gamma^{\rm p}  j_0 -(D_z/\bar{l}_{\rm sf})\delta m
=Q_{zz}.
\label{QQFM}
\end{equation}
The previous two equations, together with the first and last
(longitudinal) boundary conditions in (\ref{BC}), determine the
four interfacial quantities $\delta n_0$, $m_z$, $\delta m$, and
$Q_{zz}$ in terms of  $j_0$, $\gamma^{\rm p}$, $D/l_{\rm sf}$,
$D_z/\bar{l}_{\rm sf}$, $X_{\uparrow}$, $X_{\downarrow}$, $w_0$,
$w_{\uparrow}$, and $w_{\downarrow}$. In particular, all four are
proportional to $j_0$, and thus change sign when the current
reverses as argued earlier.

Returning to (\ref{mNM}),  we choose ${\bf H}_{\rm ext}=H\hat{\bf z}$, so the
transverse component of (\ref{LLGnm}) is solved by
\begin{equation}
m_\perp (z)=m_\perp \exp(\kappa z),
\label{mperp}
\end{equation}
where  $\kappa$ satisfies
\begin{equation}
l^2_{\rm sf}\kappa^2=1+l^2_{\rm sf}K^2+i\tau_{\rm sf}(\omega - \gamma H)
\label{dynamick}
\end{equation}
rather than the formulae quoted after (\ref{nmzz}).
Using (\ref{mperp}), the transverse component of
(\ref{spincurrent}) yields
\begin{equation}
\label{QQQNM} {\bf Q}^{\rm NM}_{\perp z}=-D\kappa{\bf m}_\perp
\label{Qperpp}
\end{equation}
while the transverse boundary conditions in (\ref{BC}) are
\begin{equation}
{\bf Q}^{\rm NM}_{\perp z}-{\bf u}_\perp Q^{\rm NM}_{zz} = w_0({\bf m}_\perp-{\bf u}_\perp m_z).
\label{QBC}
\end{equation}
From these, we can solve for the four interfacial quantities ${\bf Q}^{\rm NM}_{\perp z}$ and
${\bf m}_\perp$. The spin-transfer torque
follows straightforwardly from (\ref{Nst}) when we follow Appendix B and
express ${\bf Q}^{\rm NM}_{\perp ' z}$ in
unprimed coordinates (z-axis parallel to the interface normal) rather than
primed coordinates (z'-axis parallel to ${\bf M}$):
\begin{equation}
{{\bf N}_{\rm st} \over A}={\bf Q}^{\rm NM}_{\perp z}-{\bf u}_{\perp}Q^{\rm NM}_{zz}=
{\bf u}_{\perp}\Lambda{\kappa l_{\rm sf}-1 \over \kappa l_{\rm sf}+\Lambda}Q_{zz}.
\label{NNst}
\end{equation}
The constant $\Lambda=w_0 l_{\rm sf}/D$.
\subsection{Interfacial Exchange Torque}

The  torque produced by the surface exchange field (\ref{Hsex}) over an area $A$ at
$z=0$ is computed by integrating the torque density $-\gamma {\bf M} \times {\bf H}^{\rm S}_{\rm ex}$
over an infinitesimally thin pillbox as in (\ref{Nst}). The result is
\begin{equation}
{{\bf N}_{\rm ex} \over A}=-{\omega_0 l^2_{\rm ex} \over  m_{\rm s}}{\bf M} \times
\partial_z {\bf M}=-i{\omega_0 l^2_{\rm ex} \over m_{\rm s}} M^2(z) {\bf u'}_\perp
\label{Nex}
\end{equation}
because
\begin{equation}
\hat{\bf z} \times {\bf u}_\perp = i{\bf u}_\perp
\label{ztimesu}
\end{equation}
corresponds to a $90^\circ$ rotation of the vector ${\bf u}_\perp$. Since
the function $M(z)$ in (\ref{Mofz}) is known at this point, the only unknown in
(\ref{Nex}) is ${\bf u'}_\perp = \partial_z {\bf u}_\perp (z)|_{z=0}$.

We determine ${\bf u'}_\perp$ numerically by integrating the
transverse component of the Landau-Lifshitz-Gilbert equation
(\ref{LLGm}) inward from $z= \infty$ to $z=0$. The equation of
interest is
\begin{eqnarray}
0 & = & \left\{i\omega +\alpha \omega -i\gamma H \right\}{\bf u}_\perp \nonumber \\ \nonumber \\
&+ &
{1 \over M(z)}\left\{\textstyle{1 \over 2}\hbar\gamma^{\rm p}j_0-D_z M'(z)\right\}{\bf u'}_\perp
\nonumber \\ \nonumber \\
  & +  &
i{\omega_0 M(z) l^2_{\rm ex} \over m_{\rm s}} \left\{{\bf u}''_\perp +2{M'(z) \over M(z)}{\bf u'}_\perp
-K^2 {\bf u}_\perp \right\}.
\label{uperp}
\end{eqnarray}
The first term in brackets comes from the $\dot{M}_\alpha$,
damping, and external field terms. The second term in brackets
comes from the spin-transfer torque in the bulk of the ferromagnet
due to the spatial variation of the magnetization. The third term
in brackets is the bulk exchange torque. On account of (\ref{mm}),
the spin flip term ${\bf u}_\perp \delta m / \bar{\tau}_{\rm sf}$
in (\ref{LLGm}) cancels a piece of the spin-transfer torque. To
derive (\ref{uperp}), we have used (\ref{ztimesu}) repeatedly and
exploited the identity
\begin{equation}
\nabla^2(M{\bf u})={\bf u}\nabla^2 M +2 \nabla M \cdot \nabla {\bf u}+M \nabla^2 {\bf u}.
\label{identity}
\end{equation}

Very far from the interface, $M(z) \rightarrow m_{\rm s}$ and (\ref{uperp}) reduces to
\begin{eqnarray}
0&=&(i\omega +\alpha \omega -i \gamma H -i\omega_0K^2 l^2_{\rm ex}){\bf u}_\perp \nonumber \\
\nonumber \\
&+&\textstyle{1\over 2}\hbar\gamma^{\rm p}(j_0/m_{\rm s}){\bf u'}_\perp
+i\omega_0 l^2_{\rm ex}{\bf u''}_\perp.
\label{bulkuperp}
\end{eqnarray}
This equation is solved by ${\bf u}_\perp (z)= {\bf u}_\perp
\exp(-p\,z)$ where (${\rm Re}\, p>0)$,
\begin{eqnarray}
 p\,l_{\rm ex} =
-i{\Omega \over 2 \omega_0} \pm  {1 \over 2 \omega_0}
\sqrt{ -\Omega^2
- 4 \omega_0 \Delta },
\label{bulkmode}
\end{eqnarray}
and $ \Delta= \omega -i \omega\alpha - \gamma H -\omega_0 K^2 l_{\rm ex}^2$, where
\begin{eqnarray}
\label{bulkOmega} \Omega = {\hbar \gamma^{\rm p}  j_0 \over 2
l_{\rm ex} m_{\rm s}}.
\end{eqnarray}
This solution is used to start the inward integration of (\ref{uperp}).
We note in passing that if $\alpha=K=0$ and $q=ip$, (\ref{bulkmode}) simplifies
to
\begin{equation}
\omega(q)=\gamma H+\omega_0 l^2_{\rm ex}q^2-\Omega l_{\rm ex} q.
\end{equation}
This is (essentially) the dispersion relation derived by Bazaliy and co-workers \cite{Bazaliy:1998}
for bulk spin waves in the presence of a specified spin current.
An instability occurs ($\omega$ goes negative) when
the current is large enough that $\Omega^2 > 4 \omega_0 \gamma H$.  If we take $\mu_0 H =1~{\rm T}$,
this gives a charge current density  $J_0=|e|j_0 > 10\times 10^9$~A/cm$^2$, which
is an order of magnitude larger than
the interfacial instabilities we discuss below.

It is now straightforward to search for precession-type instabilities. For given values of
current $j_0$, external magnetic field $H$, and transverse wave vector $K$, we search
through complex frequency ($\omega=\omega_{\rm  R}+i\omega_{\rm I}$) space looking for modes. That is,
situations where (\ref{zerotorque}) is satisfied using (\ref{NNst}) for the interfacial spin-transfer
torque and (\ref{Nex}) for the interfacial exchange torque. If $\omega_{\rm I}<0$,
the mode is linearly unstable against precession.
A suitable initial guess is the zero-current surface spin wave frequency
\begin{equation}
\omega_{\rm s}={\gamma H + \omega_0 l_{\rm ex}^2 K^2 \over 1-i\alpha}.
\end{equation}

\section{Thin Film Geometry}
\label{asymint}
 As in the previous section, the search for
current-induced precessional instabilities in the thin film
geometry of Figure~\ref{fig:geometry}(b) begins by solving the
longitudinal part of the problem. The spin accumulation decays
exponentially into the non-magnetic leads so at the left ($z=0$)
and right ($z=t$) interfaces, we have the analog of (\ref{QQNM})
for the interfacial spin currents:
\begin{equation}
Q^{\rm L}_{zz}=-{D^{\rm L}_{\rm s} \over l^{\rm L}_{\rm sf}}m^{\rm L}_z \hspace{30pt}
 Q^{\rm R}_{zz}=-{D^{\rm R}_{\rm s} \over l^{\rm R}_{\rm sf}}m^{\rm R}_z
\label{twoQzzNM}
\end{equation}
The superscript on the diffusion constants and on the spin flip lengths allows us to
assign different material properties to the leads and thus to the interfaces. \cite{Polianski:2003}

The $z$-dependences of the spin accumulation and the spin current in the ferromagnetic film
are linear combinations of $\exp(\pm z/\bar{l}_{\rm sf})$. Straightforward algebra yields the interfacial relations
\begin{eqnarray}
Q^{\rm L}_{zz}&=&\gamma^{\rm p}j_0   \nonumber \\
& & \mbox{} - {D_z \over \bar{l}_{\rm sf}}
\left[\delta m^{\rm L}\coth (t /  \bar{l}_{\rm sf})
-\delta m^{\rm R}{\rm csch}(t / \bar{l}_{\rm sf})\right] \nonumber  \\
Q^{\rm R}_{zz}&=&\gamma^{\rm p}j_0  \nonumber \\
& & \mbox{} - {D_z \over \bar{l}_{\rm sf}}
\left[-\delta m^{\rm R}\coth (t /  \bar{l}_{\rm sf})
+\delta m^{\rm L}{\rm csch}(t / \bar{l}_{\rm sf}) \right]. \nonumber  \\
\label{twoQzzFM}
\end{eqnarray}
The four equations in (\ref{twoQzzNM}) and (\ref{twoQzzFM}), together with the two longitudinal boundary
conditions in (\ref{BC}) at each interface, determine the eight quantities $\delta n_0^{\rm L}$,
$\delta n_0^{\rm R}$, $m^{\rm L}_z$, $m^{\rm R}_z$, $\delta m^{\rm R}$, $\delta m^{\rm L}$,
$Q^{\rm L}_{zz}$, and $Q^{\rm R}_{zz}$ in terms of the physical parameters of the model.
Finally, because it is used in the integration of (\ref{uperp}), we record that
\begin{eqnarray}
M(z) &=& m_{\rm s} + \delta m^{\rm L} {\rm csch}(t/\bar{l}_{\rm sf})\sinh[(t-z)/\bar{l}_{\rm sf}] \nonumber \\
& & \mbox{} + \delta m^{\rm R} {\rm csch}(t/\bar{l}_{\rm sf})\sinh(z/\bar{l}_{\rm sf}).
\label{filmspinacc}
\end{eqnarray}

Up to an overall sign, the spin transfer and exchange torques are still given by
(\ref{NNst}) and (\ref{Nex}). Therefore, at the right interface,
 we choose ${\bf u}_\perp$ arbitrarily
and compute ${\bf u'}_\perp$ so that (\ref{zerotorque}) is
satisfied at that interface. This provides the initial condition
for the integration of (\ref{uperp}) toward $z=0$. Then, the
search through complex frequency space for solutions that satisfy
(\ref{zerotorque}) at the left interface proceeds as described
earlier. As initial guesses, we use the zero-current frequencies
of the spin wave modes of the film:
\begin{equation}
\label{modefreq}
 \omega_n={\gamma H + \omega_0 l_{\rm ex}^2 [(n
\pi/t)^2 +K^2] \over 1-i\alpha}.
\end{equation}

\section{Results \& Discussion}

The numerical results presented in this section are based on
converged solutions of the transport and magnetization equations
using material parameters derived directly from experiment or from
first-principle electronic structure calculations
(Table~\ref{tab:parameters}).

\begin{table}
\begin{tabular}{lrlr}
Parameter & \multicolumn{2}{c} {Value} & Ref. \\ \hline
$\rho ({\rm Cu})$ & 6 & n$\Omega$m & \onlinecite{Bass:1999} \\
$\rho_{\uparrow}$ & 40.5 & n$\Omega$m & \onlinecite{Bass:1999} \\
$\rho_{\downarrow}$ & 109.5 & n$\Omega$m & \onlinecite{Bass:1999} \\
$l_{\rm sf}({\rm Cu})$ & 450 & nm & \onlinecite{Yang:1994} \\
$\bar{l}_{\rm sf}({\rm Co})$ & 59 & nm & \onlinecite{Piraux:1998} \\
$AR_\uparrow$ & 0.117 & f$\Omega$m$^2$ & \onlinecite{Bass:1999} \\
$AR_\downarrow$ & 0.903 & f$\Omega$m$^2$ & \onlinecite{Bass:1999} \\
$A_{\rm FS}$ & 580 & nm$^{-2}$ & \onlinecite{AshcroftMermin}  \\
$\partial n/\partial\mu$ & 6.23 & eV$^{-1}$nm$^{-3}$ & \onlinecite{papa} \\
$\partial n_\uparrow/\partial\mu_\uparrow$ & 8.17 & eV$^{-1}$nm$^{-3}$ & \onlinecite{papa} \\
$\partial n_\downarrow/\partial\mu_\downarrow$ & 31.6 & eV$^{-1}$nm$^{-3}$ & \onlinecite{papa} \\
$A_{\rm ex}$ & $2 \times 10^{-11}$ & J/m & \onlinecite{exchange} \\
$M_{\rm s}$ & $14.46\times 10^5$ & A/m & \onlinecite{saturationmag} \\
$\gamma ({\rm Co})$ & 1.9$\times 10^{11}$ & s$^{-1}$T$^{-1}$ & \onlinecite{gammaCo} \\
$\gamma ({\rm Cu})$ & 1.75$\times 10^{11}$ & s$^{-1}$T$^{-1}$ & \onlinecite{AshcroftMermin} \\
$\alpha$ & 0.01 & & \onlinecite{McMichael}\\
\hline
$D$ & $4.1\times 10^{16}$ & nm$^2$s$^{-1}$ & \\
$D_z$ & $1.7\times 10^{15}$ & nm$^2$s$^{-1}$ & \\
$\tau_{\rm sf}({\rm Cu})$ & $2.4 \times 10^{-11}$ & s & \\
$\bar{\tau}_{\rm sf}({\rm Co})$ & $2.0 \times 10^{-12}$ & s & \\
$\gamma^{\rm p}$ & 0.46 &  &  \\
$\omega_{0}$ & 3.465 $\times 10^{11}$ & s$^{-1}$ & \\
$l_{\rm ex}$ & 3.9 & nm & \\
$2 m_{\rm s}/\hbar$ & 143.6 & nm$^{-3}$ &  \\
$w_0$ & $11.4\times 10^{14}$ & nms$^{-1}$  &  \\
$w_\uparrow$ & $8.7\times 10^{14}$ & nms$^{-1}$ & \\
$w_\downarrow$ & $2.25\times 10^{14}$ & nms$^{-1}$ & \\
$X_\uparrow$ & 1.90 &  &  \\
$X_\downarrow$ & 0.24  &  &  \\
\end{tabular}
\caption{Material parameters for Cu/Co. The values above the
horizontal line were taken from experiment or first-principles
calculation. The values below the line were calculated using
equations given in the text, $\sigma=1/\rho$, and the Einstein
relation $e^2 \sigma = D
\partial n/\partial \mu$.} \label{tab:parameters}
\end{table}

\subsection{Single Interface Geometry}

\begin{figure}
\includegraphics{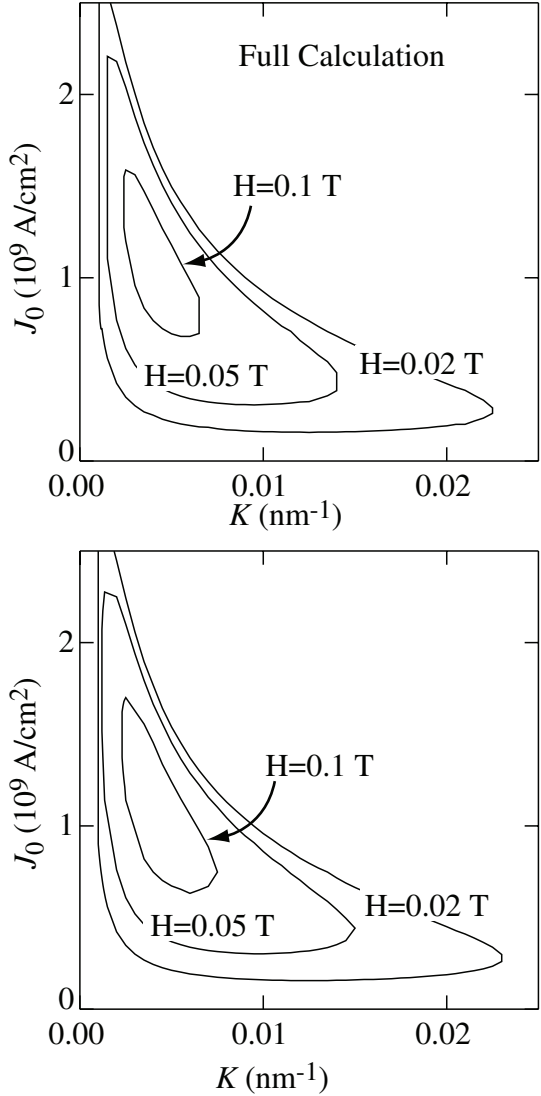}
\caption{Regions of instability for the single interface geometry
of Figure~\ref{fig:geometry}(a). Each contour gives the region in
$J_0$-$K$ space within which instabilities occur for an series of
applied fields (labeled on each contour). The top panel gives the
results for the full calculation in the ferromagnet.  The bottom
panel gives the results if the effect of spin accumulation on the
exchange interaction is ignored.
} \label{fig:stability}
\end{figure}

The top panel of Figure~\ref{fig:stability} illustrates the
regions of linear instability as a function of applied charge
current $J_0$ and wave vector $K$ parallel to the interface for
the single interface geometry (top panel of
Figure~\ref{fig:geometry}). An initial state of current induced
precession occurs inside each contour labeled by a choice of
external magnetic field $H$. We emphasize that every contour forms
a closed loop in the $J_0-K$ plane. That is, for a fixed value of
$J_0$, the instability region is bounded by a lower and an upper
critical wave vector and, for a fixed value of $K$, the
instability region is bounded by a lower and an upper critical
current density. Note also that the instability occurs only for
rather small values of external magnetic field. We will see in the
next subsection that thin films support precession for much larger
values of $H$.  In addition, it appears that the upper critical
field is very sensitive to the parameters that describe the
system.  A systematic exploration of parameter space is
prohibitive, but we note that the choices $l_{\rm sf}=1000$~nm and
$\overline{l}_{\rm sf}=15$~nm give regions of instability for
$H>2$~T.

The bottom panel of Figure~\ref{fig:stability} is the same as the top panel except that
we have replaced
the full transverse Landau-Lifshitz-Gilbert equation (\ref{uperp}) by
\begin{eqnarray}
0 & = & \left\{i\omega +\alpha \omega -i\gamma H \right\}{\bf u}_\perp \nonumber \\ \nonumber \\
&+ &
{1\over m_{\rm s}}\left\{\textstyle{1 \over 2}\hbar\gamma^{\rm p}j_0-D_z M'(z)\right\}{\bf u'}_\perp
\nonumber \\ \nonumber \\
  & +  &
i\omega_0 l^2_{\rm ex}  \left\{{\bf u}''_\perp
-K^2 {\bf u}_\perp \right\}.
\label{uperpsimple}
\end{eqnarray}
This approximation neglects the ferromagnetic spin accumulation $\delta m(z)$
in the exchange interaction and retains only its gradient
variation in the bulk spin-transfer torque.  Evidently, these omissions make no
difference to the qualitative features of the diagram.

According to Figure~\ref{fig:stability}, no precession occurs if
either the current density, the magnetic field, or the parallel
wave vector gets too large. We expect the instability to disappear
when the Gilbert damping overwhelms the effect of the spin
transfer torque. Therefore, since the damping parameter $\alpha$
enters the theory only in the combination $\alpha \omega$ [see
(\ref{uperpsimple})], we anticipate that the mode frequency is an
increasing function of $j$, $H$, and $K$.

To check this, we approximate the spin current in the second line
of (\ref{uperpsimple}) by its value at the interface to get
\begin{eqnarray}
0 & = & \left\{i\omega +\alpha \omega -i\gamma H \right\}{\bf u}_\perp
\nonumber \\ \nonumber \\
&+ & {1\over m_{\rm s}}\left\{\textstyle{1 \over
2}\hbar\gamma^{\rm p}j_0-D_z \delta m(0)/\bar{l}_{\rm
sf}\right\}{\bf u'}_\perp
\nonumber \\ \nonumber \\
  & +  &
i\omega_0 l^2_{\rm ex}  \left\{{\bf u}''_\perp -K^2 {\bf u}_\perp
\right\}. \label{uperpssimple}
\end{eqnarray}
This linear equation, like (\ref{bulkuperp}), is solved by ${\bf
u}_\perp (z)= {\bf u}_\perp \exp(-p\,z)$. Therefore,
\begin{equation}
\label{oneratio} {u'_\perp(0) \over u_\perp (0)}=-p,
\end{equation}
with  $p$ given by (\ref{bulkmode}) but with (\ref{bulkOmega})
replaced by
\begin{equation}
\label{2Omega} \Omega={\hbar\gamma^{\rm p}j_0/2-D_z \delta
m(0)/\bar{l}_{\rm sf} \over l_{\rm ex}m_s} ={\hbar\gamma'^{\rm
p}j_0 \over 2l_{\rm ex} m_s}.
\end{equation}
On the other hand, replacing $\kappa$ by $K$ in (\ref{NNst})
(large $K$ approximation) and $M(0)$ by $m_s$ in (\ref{Nex}), the
mode condition (\ref{zerotorque}) gives
\begin{equation}
\label{tworatio} {u'_\perp(0) \over u_\perp (0)}=-i{Q_{ zz}\Lambda
\over \omega_0 l^2_{\rm ex}m_s}{Kl_{\rm sf}-1 \over Kl_{\rm
sf}+\Lambda} =-i{Q_{zz}\Lambda' \over \omega_0 l^2_{\rm ex}m_s}.
\end{equation}
Equating (\ref{oneratio}) to (\ref{tworatio}) gives an estimate of
the spin wave mode frequency $\omega=\omega_{\rm R} +i \omega_{\rm
I}$. We find
\begin{eqnarray}
\label{omegaestimate}
\omega_{\rm I}&=&\alpha \omega_{\rm R} \nonumber\\
\omega_{\rm R}&=& \left\{{j_0^2 \over \omega_0 l_{\rm ex}^2
m_s^2}\left[\gamma'^{\rm p}\Lambda'\left({Q_{zz}\over j_0}\right)
+\Lambda'^2\left({Q_{zz}\over j_0}\right)^2 \right] \right.
\nonumber \\ &+&\omega_0 l_{ex}^2 K^2+\gamma H \left. \right\}
\times {1 \over 1 + \alpha^2}.
\end{eqnarray}

The second line in (\ref{omegaestimate}) confirms our expectation
about the behavior of the mode frequencies and thus rationalizes
the topology of Figure~\ref{fig:stability}. Moreover, this
analytic formula for $\omega_{\rm R}$ agrees {\it quantitatively}
with the real part of the mode frequencies we find from our
numerical solution of the full equations of motion. Unfortunately,
(\ref{omegaestimate}) also predicts $\omega_{\rm I}>0$. That is,
(\ref{uperpssimple}) predicts only stable spin waves. These modes
are unstable in the full solution of (\ref{uperpsimple}), but only
barely so ($\omega_{\rm I}$ is negative but very small in
magnitude).  In fact, the stability or instability of a given mode
turns out to be an extremely sensitive function of the phase of
$u'_\perp(0)/u_\perp (0)$ and the factor $M'(z)$ in
(\ref{uperpsimple}). The latter is determined by the ferromagnetic
spin flip length $\bar{l}_{\rm sf}$. We conclude that variation of
the magnetization with the longitudinal variable $z$ is an {\it
essential} ingredient for instability to occur in the
single-interface geometry.

\begin{figure}
\includegraphics{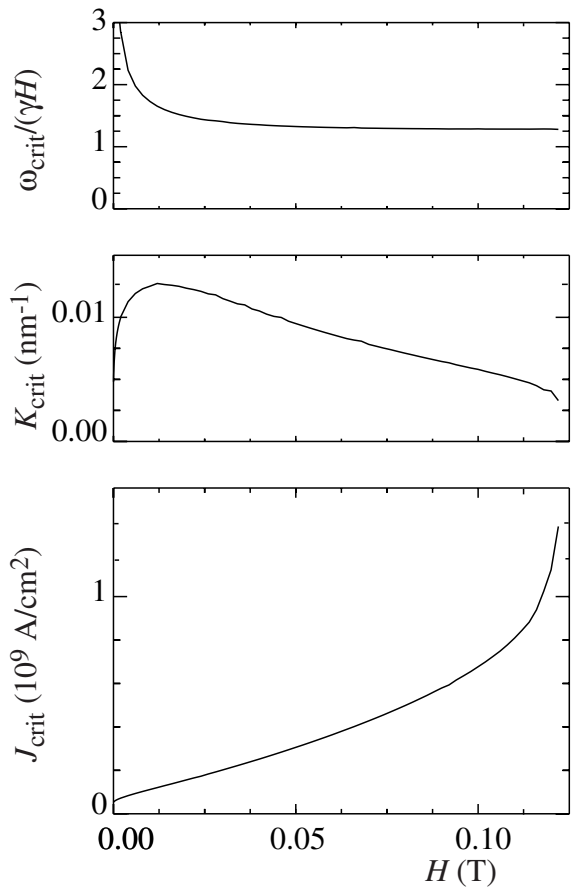}
\caption{Dependence on external magnetic field $H$ of the critical
precession frequency $\omega_{\rm crit}$ divided by the Larmor
frequency (top panel); the critical parallel wave vector $K_{\rm
crit}$ (middle panel); and the critical current density $J_{\rm
crit}$ (bottom panel) for the single interface geometry of
Figure~\ref{fig:geometry}(a).
} \label{fig:jcrit}
\end{figure}

Figure~\ref{fig:jcrit} shows the calculated field dependence of
the critical frequency $\omega_{\rm crit}$, the critical
transverse wave vector $K_{\rm crit}$, and the critical current
density $J_{\rm crit}$,  where precession first occurs in the
single interface case. Consistent with (\ref{omegaestimate}), all
three are finite as $H\to0$. Otherwise, $\omega_{\rm crit}$ and
$J_{\rm crit}$ vary roughly linearly with field, as they do in
experiments performed on single films, \cite{Ji:2003} until the
instability shuts off for large enough $H$. The order of magnitude
of the critical current $(\approx 10^9$ A/cm$^{\rm 2}$) is also
consistent with the single film experiments (and roughly ten times
larger than the critical current seen in the trilayer pillars
studied in Ref.~\onlinecite{Myers:1999}). Note, finally, that
$\omega_{\rm crit}$ deviates systematically from the Larmor
frequency $\gamma H$.

\subsection{Thin Film Geometry}

We remarked at the end of Section~\ref{sec:origin} that the
interfacial spin torque associated with current flow from a
ferromagnet to a non-magnet is opposite in direction to the
interfacial spin torque associated with current flow from a
non-magnet to a ferromagnet. As pointed out by Polianski and
Brouwer, \cite{Polianski:2003} this implies that {\it zero} net
spin torque acts on a {\it uniformly} magnetized film attached to
identical leads at identical interfaces. Those authors used leads
of different lengths to break the mirror symmetry of this
geometry. We discuss that case in more detail in the next section.

In our calculations for a ferromagnetic film sandwiched between
two semi-infinite, non-magnetic leads (middle panel of
Figure~\ref{fig:geometry}), the natural spatial variation of the
magnetization of the spin wave modes breaks the longitudinal
symmetry.  Additional spatial variation arises because there is
current-induced mixing amongst the modes (Figure~\ref{fig:modes}).
As in the single-interface problem, the spatial variation $M(z)$
is essential for precession {\it if} we assume a symmetrical
geometry where the scattering properties of the two interfaces are
identical. We do this here, even though the formulae given in
Section~\ref{asymint} allow for a more general, asymmetrical
situation.

Figure~\ref{fig:stability_thick} illustrates precessional
instability diagrams for a ferromagnetic film of thickness
$t=40$~nm. For each value of external magnetic field, the integer
that labels each tear drop shaped region indicates the mode that
goes unstable in that region.
Notice that the area of the envelope in the $J_0-K$ plane that
encloses all the $H=0.05$~T teardrops is much larger than the area
of the $H=0.05$~T instability region for the single-interface
problem sketched in Figure~\ref{fig:stability}. The spin transfer
torque is roughly the same in the two cases, but it is much easier
to excite spin waves in the film because there is much less
ferromagnetic ``mass'' present to contribute to the damping.

The field dependence of the critical current $J_{\rm crit}$ is
shown in Figure~\ref{fig:jcrit_h_finite} for several film
thicknesses.  A relatively large current density is needed to
excite the 10 nm film because the symmetry breaking required to
generate a net torque occurs only when the lowest frequency spin
wave mode (which has uniform magnetization in the $z$-direction)
is perturbed by current-induced mixing with the next-highest spin
wave mode (which has non-uniform magnetization in the
$z$-direction). This is shown by the leftmost panel of
Figure~\ref{fig:modes}. We will use (\ref{2Omega}) as the
characteristic frequency associated with the driving current and
(\ref{modefreq}) to estimate the frequency splitting between the
$n=1$ and $n=2$ film modes. Setting these equal, we get a
field-independent estimate of $J_{\rm crit}$ of the order of
$10^9-10^{10}$ ${\rm A/m}^2$.

\begin{figure}
\includegraphics{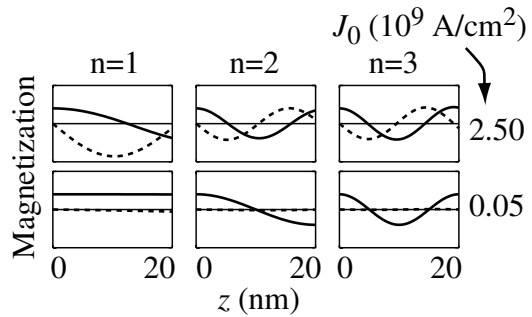}
\caption{Current dependence of the spatial variation of the
magnetization for the three lowest spin wave modes of the
ferromagnetic thin film shown in the middle panel of
Figure~\ref{fig:geometry}. Solid curves are $M_{\rm x}(z)$. Dashed
curves are $M_{\rm y}(z)$. There is almost no mode mixing for
$J_0=0.05 \times 10^9$ A/cm$^2$.}\label{fig:modes}
\end{figure}

The critical current density decreases in
Figure~\ref{fig:jcrit_h_finite} as the film thickness increases to
20 nm and then to 40 nm because the mode splitting decreases in
typical particle-in-a-box fashion. We will refer to this below as
a quantum size effect. Then, beginning with the 40 nm film, a new
phenomenon occurs. For large enough $H$, the perturbed $n=1$
precession mode {\it never} goes unstable (there is no tear drop
labelled ``$1$'' in the $H=$~2~T panel of
Figure~\ref{fig:stability_thick}). Instead, an upward tick in
$J_{\rm crit}$ occurs when the perturbed $n=2$ spin wave mode
begins to precess. For the 80 nm film, this phenomenon moves to
lower values of $H$ and successively higher spin wave modes become
the first to become unstable. This transition of instability from
one mode to the next highest is responsible for the jagged
appearance of the $J_{\rm crit}(H)$ curves for the thicker films.

\begin{figure}
\includegraphics{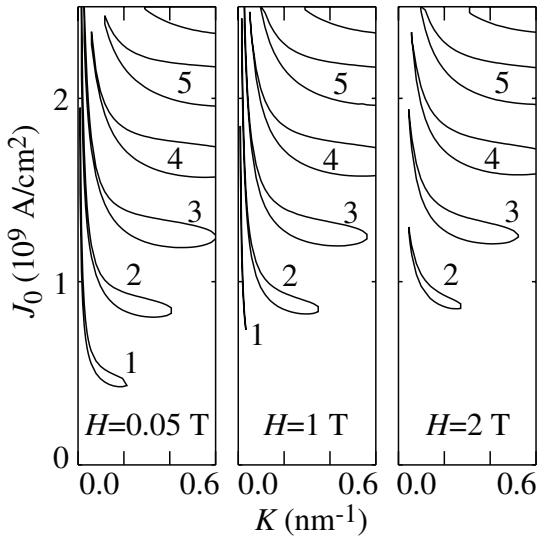}
\caption{Instability diagram for a $t=40$ nm ferromagnetic film in
the geometry of Figure~\ref{fig:geometry}(b) for three values of
external magnetic field. The integers label the thin film
precessional mode that goes unstable in each region.}
\label{fig:stability_thick}
\end{figure}

The nearly vertical line segment in
Figure~\ref{fig:jcrit_h_finite} is the \mbox{$t=\infty$}
single-interface result redrawn from the lowest panel of
Figure~\ref{fig:jcrit}.  It is smooth because the region of
instability for each mode becomes increasing small (as a function
of $H$) until, in the limit of a semi-infinite film, every value
of $H$ leads to the linear instability of a different precession
mode of the film. This curve terminates at a critical field
$H_{\rm crit}(t=\infty)$. The 120 nm film curve shows that $H_{\rm
crit}(t)$ increases as the film thickness decreases from
$t=\infty$.

Finally, Figure~\ref{fig:jcrit_thick} summarizes the thickness
dependence of $J_{\rm crit}$ for three values of external magnetic
field. For $H=0.1$~T, the critical current decreases monotonically
as $t$ increases as the quantum size effect argument suggests.
However, to reach the $t\to\infty$ limit (horizontal arrow),
$J_{\rm crit}$ must eventually increase. As the $H=1\, {\rm T}$
and $H=3\,{\rm T}$ curves show, this occurs in jumps each time a
mode becomes absolutely stable and the next highest mode initiates
the precession. $J_{\rm crit}(t)$ oscillates in this regime
because the quantum size effect operates as long as one particular
mode is responsible for the instability.

The $H=1\, {\rm T}$ and $H=3\,{\rm T}$ curves in
Figure~\ref{fig:jcrit_thick} terminate at a critical thickness
$t_{\rm crit}(H)$ that decreases with increasing magnetic field.
This is consistent with the fact that the semi-infinite
ferromagnet only supports current-induced precession for
relatively small values of external magnetic field.

\begin{figure}
\includegraphics{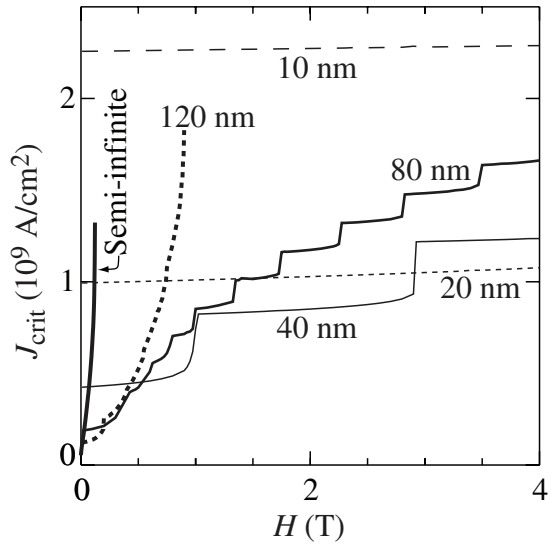}
\caption{Critical current as a function of field for several
thicknesses of the ferromagnetic film in the geometry of
Figure~\ref{fig:geometry}(b). The non-smooth behavior of the
curves for the thicker films is associated with transitions
between instabilities associated with different thin film modes. }
\label{fig:jcrit_h_finite}
\end{figure}

\begin{figure}
\includegraphics{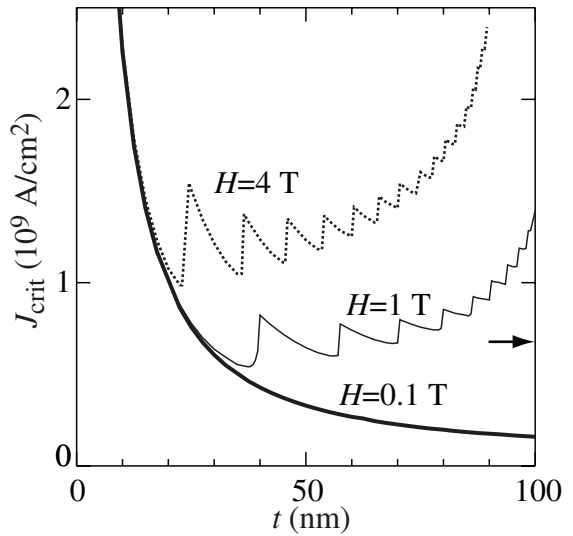}
\caption{Critical current as a function of ferromagnetic film
thickness for three values of external magnetic field in the
geometry of Figure~\ref{fig:geometry}(b). A jump occurs whenever a
new mode is responsible for the precession. The horizontal arrow
is the $H=0.1$~T critical current for a semi-infinite
ferromagnetic film. } \label{fig:jcrit_thick}
\end{figure}

\section{Effect of Finite Leads}

Polianski and Brouwer\cite{Polianski:2003} studied current-induced
magnetization precession for the geometry depicted in the lowest
panel of Figure~\ref{fig:geometry}. The unequal  length leads that
break the symmetry of the problem are each attached to an electron
``reservoir''. By definition, the electrochemical potential takes
its equilibrium value inside each reservoir. Therefore, the spin
accumulation (a non-equilibrium property) must go smoothly to zero
when we enter each reservoir from the adjacent lead. These authors
also fixed the magnetization of the ferromagnetic film to be
uniform in the $z$ direction.

In this section, we study this reservoir geometry using our
formalism, both with and without the additional restriction to
uniform longitudinal magnetization in the ferromagnet.  Our
expectation is that the approximations of Ref.~\onlinecite{Polianski:2003}
will be adequate if the ferromagnetic film thickness $t$ is less
than the exchange length $l_{\rm ex}$.

Let the non-magnetic leads have lengths $t^{\rm L}$ and $t^{\rm
R}$ as shown in Figure~\ref{fig:geometry}(c). It is
straightforward to check that the solution of (\ref{LLGnm}) in,
say, the right lead that satisfies the reservoir boundary
condition of zero longitudinal and transverse spin accumulation at
$z=t+t^{\rm R}$ gives at the interface
\begin{eqnarray}
m_z(t)  &=& m_z^{+} \left( \exp [ -t^{\rm R}/{l}_{\rm sf}]
                         - \exp [  t^{\rm R}/{l}_{\rm sf}]
   \right) \nonumber\\\nonumber \\
m_\perp(t)  &=& m_\perp^{+} \left( \exp [-\kappa t^{\rm R}]
                                 - \exp [ \kappa t^{\rm R}]
                                 \right).
\end{eqnarray}
The corresponding spin current (\ref{spincurrent}) is
\begin{eqnarray}
Q_{zz}(t)  &=& ( D m_z^{+} / {l}_{\rm sf} )
   \left( \exp [ -t^{\rm R}/{l}_{\rm sf}]
        + \exp [  t^{\rm R}/{l}_{\rm sf}] \right)
        \nonumber\\\nonumber \\
Q_{\perp z} (t)  &=& ( \kappa D m_\perp^{+} )
   \left( \exp [-\kappa t^{\rm R}] + \exp [ \kappa t^{\rm R}] \right) .
\end{eqnarray}
These can be recast as a relation between the spin accumulation
and the spin current at the interface:
\begin{eqnarray}
Q_{zz}(t)  &=& -{ D  \over {l}_{\rm sf} } m_z(t)
\coth(t^{\rm R}/{l}_{\rm sf})
 \nonumber\\\nonumber \\
Q_{\perp z} (t)  &=& - \kappa D  m_\perp(t)
\coth(\kappa t^{\rm R}). \label{coth}
\end{eqnarray}
These formulae are the same as those used previously except for
multiplicative factors depending on the length of the leads. We
recover (\ref{QQNM}) and (\ref{QQQNM}) in the limit $t^{\rm
R}\rightarrow\infty$. Very similar results apply to the left lead
and interface.

The modified spin currents (\ref{coth}) can be inserted into the
solution of the longitudinal spin accumulation and then the
boundary conditions for the interfacial spin torque. Wherever $D$
appears explicitly (i.e. ${l}_{\rm sf}$ and $\kappa$ are not
affected) it can be replaced by $D$ times the factor given above.
Otherwise, the full calculation for the reservoir geometry
proceeds as outlined in previous sections.
Figure~\ref{fig:asymmetric} illustrates results for two films with
$t^{\rm L}=\infty$. The $t^{\rm R}=450$~nm is {\it weakly lead
asymmetric} in the sense that the factor $\coth(t^{\rm R}/{l}_{\rm
sf})$ in (\ref{coth}) takes the value $1.3$. The $t^{\rm R}=45$~nm
film is {\it strongly lead asymmetric} because $\coth(t^{\rm
R}/{l}_{\rm sf})$ has the value $10.$

\begin{figure}
\includegraphics{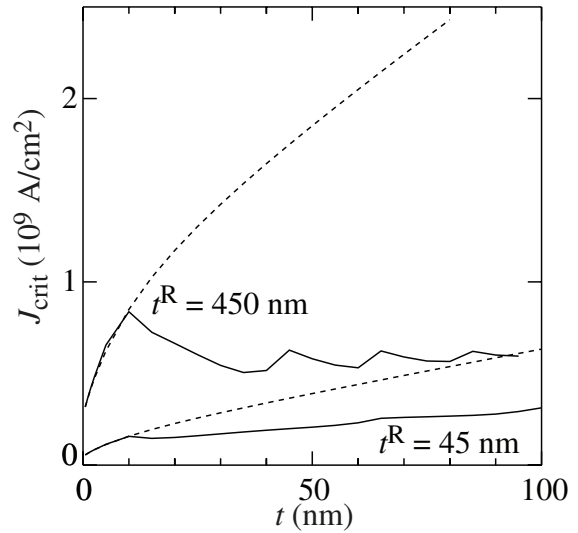}
\caption{Critical current as a function of ferromagnetic film
thickness in the asymmetrical geometry of
Figure~\ref{fig:geometry}(c) with $t^{\rm L}=\infty$. The dashed
curves are a uniform magnetization approximation. The solid curves
are the results of the full calculation. The breaks in the solid
curves correspond to the onset of precession from successively
higher film modes. The external field is
$H=3~T$.}\label{fig:asymmetric}
\end{figure}

From relatively  small values near $t=0$ (see below), $J_{\rm
crit}$ in Figure~\ref{fig:asymmetric} increases initially as the
film thickens due to increased damping from the additional
ferromagnetic material. This trend continues until the lowest mode
of uniform magnetization becomes absolutely stable and the next
higher spin wave mode initiates precession. Thereafter, the $t$
dependence of the critical current density is determined by a
competition between the quantum size effect (which tends to
decrease $J_{\rm crit}$ at $t$ increases) and increased damping
(which tends to increase $J_{\rm crit}$ as $t$ increases).

A comparison of Figure~\ref{fig:jcrit_thick} with
Figure~\ref{fig:asymmetric} shows that the very thin film limit of
$J_{\rm crit}$ depends very sensitively on the asymmetry of the
leads. When the leads have equal length
(Figure~\ref{fig:jcrit_thick}), the spin current and torque come
exclusively from the $D_zM'(z)$ term in (\ref{uperp}). This
implies a large critical current density because a large amount of
energy is needed to mix the modes when $t\to 0$ [see
(\ref{modefreq})]. By contrast, the critical current density is
relatively small when the leads have unequal length
(Figure~\ref{fig:asymmetric}) because (even if the magnetization
is uniform) there is incomplete cancellation between the
(oppositely signed) spin transfer torques generated at the two
ends of the ferromagnet.\cite{Polianski:2003}
Figure~\ref{fig:leadlength} shows that $J_{\rm crit}$ actually
increases {\it exponentially} with $t^{\rm R}$ when $t^{\rm
L}=\infty$.

\begin{figure}
\includegraphics{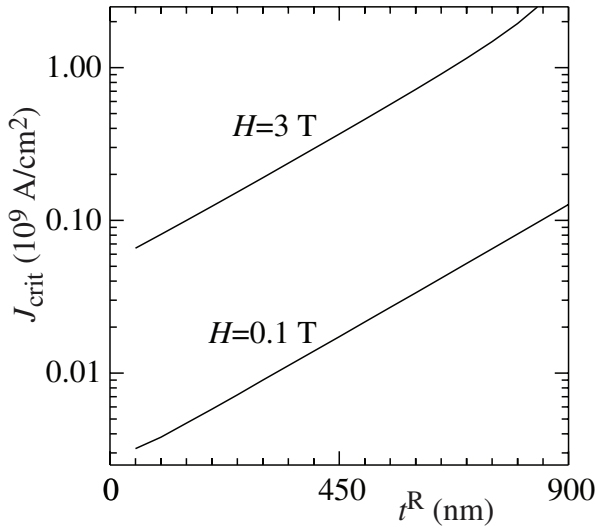}
\caption{Log-linear plot of critical current in the geometry of
Figure~\ref{fig:geometry}(c) as a function of $t^{\rm R}$ when
$t^{\rm L}=\infty$. The thickness of the ferromagnetic film is
$t=2.5\,$ nm.}\label{fig:leadlength}
\end{figure}

We now consider an approximation to these full calculations that
forces the magnetization in the ferromagnetic film to be uniform
in the $z$ direction. This occurs automatically in our numerical
work if the exchange length $l_{\rm ex}$ and the spin diffusion
length $\bar{l}_{\rm sf}$ are made much larger than any thickness
$t$ of interest. The results are shown as dashed curves in
Figure~\ref{fig:asymmetric}. As expected, the full and approximate
calculations agree quantitatively only when $t$ is smaller than or
approximately equal to the true Co exchange length of $3.9\,$nm
(see Table~\ref{tab:parameters}). The agreement for $t=10$~nm is
somewhat accidental because (as the break in the curve indicates)
the precession at this thickness is {\it not} associated with a
state of uniform magnetization in the longitudinal direction.

On the other hand, Figure \ref{fig:asymmetric} shows that for our
choice of parameters, the uniform magnetization approximation gets
relatively better (for $t\gg l_{\rm ex}$) as the lengths of the
two leads become more dissimilar. Qualitatively, this occurs
because there are two sources of net spin transfer torque in this
problem: the longitudinal asymmetry of the leads and the
longitudinal asymmetry of the magnetization. If the first
dominates the second (strong lead asymmetry), it does not really
matter which mode goes unstable and the uniform magnetization
approximation will be acceptable. If the second dominates the
second (weak lead asymmetry), the mode identity is crucial and the
approximate calculation fails badly.

\section{Limitations \& Future Directions}

In experiment, the large current densities required to produce
current-induced precession are achieved using point-contact,
nanowire, or nanopillar geometries where conductor dimensions in
the direction perpendicular to the current flow are
``necked-down'' from micron-scale dimensions to hundred
nanometer-scale dimensions. The reservoir geometry of
Figure~\ref{fig:geometry}(c) is an approximate way to model this
behavior within the context of a strictly one-dimensional
calculation. Nonetheless, it remains true that we have ignored the
fact that the current only acts over a small area. The small area
is either a small part of a large film, or a small finite element.
In both cases, we have ignored the behavior of the current at the
edges.

For a point contact geometry, we  have ignored how the current
spreads into the film, how the variation in the magnetization
decays into the film both laterally and in the interface
direction, and how these interact.  For a nanowire or nanopillar
geometry, we have  ignored the behavior of the current at the
edges. Clearly, these three-dimensional effects will need to be
considered if calculations of the sort presented here are ever to
be truly predictive in the sense of device design.

Our calculations take no account of magnetostatics. The necessary
computations are a bit complex because the magnetic self-energy is
a non-local function of the spin density. It is true that the
dominant effect can be modelled as magnetic anisotropy (which we
have also ignored), but the known effects of magnetostatics on
spin-wave dispersion suggest the possibility of other effects for
the present problem.

More significantly, we have focused exclusively on the initial,
linear instability of a current-carrying nano-magnet toward
precession. In reality, non-linear effects cut off the exponential
growth of the mode amplitudes we find. This means that the final
magnetic dynamical state may have little in common with simple
precession, particularly for magnets whose size significantly
exceeds an exchange length. To our knowledge, little or nothing is
known about spatially inhomogeneous magnetic states of this type
in the present context.

Finally, we have performed all of our calculations at zero
temperature. An extension to finite temperature clearly would be
interesting in light of recent experiments on thermally activated
magnetic reversal induced by a spin-polarized
current.\cite{Myers:2002}

\section{Summary}
We have combined a drift-diffusion description of transport with a
Landau-Lifshitz-Gilbert description of magnetization dynamics to
demonstrate that unpolarized current flow from a non-magnet into a
ferromagnet can produce a precession-type instability of the
magnetization. Neglecting magnetic anisotropy and magnetostatics,
our numerical calculations show (for both single-interface and
thin film geometries) that a linear instability occurs when both
the current density and the excitation wave vector parallel to the
interface are neither too small nor too large. The critical
current density generally increases with increasing magnetic field
until the instability is quenched. Compared to the
single-interface geometry, the film geometry supports a
precessional instability over a much larger range of external
magnetic field. In general, a particular spin wave mode initiates
the instability when the mode just below it (in energy) becomes
absolutely stable against current-induced precession.

In a lead-film-lead geometry, it is necessary to break mirror
symmetry to generate precession because equal and opposite spin
transfer torques are generated at the two ends of the
ferromagnetic film. In our model, the required symmetry breaking
is provided by spin wave modes of the film whose magnetization
varies along the direction of current flow. If the magnetization
is artificially constrained to be longitudinally uniform,
asymmetry must be introduced in the length or properties of the
leads or in the scattering properties of the two interfaces. When
the spin wave mode asymmetry generates the spin-transfer torque,
the critical current density first decreases (a quantum size
effect) and then slowly increases (as successive modes become
stable against precession) as the film thickness increases. When
interface or lead asymmetry dominates the spin transfer torque,
the critical current is small (compared to the mode-dominated
regime) and slowly increases due to increased damping as the film
thickness increases.

\section{Acknowledgment}

One of us (J.X.) acknowledges support from the National Science
Foundation under grant DMR-9820230.  The authors thank Dr. Robert
McMichael for essential help with the references.\\
\appendix{\bf Appendix A: Exchange Magnetic Field} \\

The total exchange energy of a semi-infinite ferromagnet that extends from
$z=z_{\rm L}$ to $z=z_{\rm R}$ is
\begin{eqnarray}
E_{\rm ex} &=& { A_{\rm ex} \over m_{\rm s}^2 }
\int d^2 R \int_{z_{\rm L}}^{z_{\rm R}} dz \,
(\partial_\alpha{\bf M})(\partial_\alpha{\bf M}).
\label{exenergy}
\end{eqnarray}
The quantity $m_{\rm s}$ occurs here because $M$ denotes spin density
rather than magnetization.
Focus on the $\alpha=z$ component and let
${\bf M} \rightarrow {\bf M} + \bm\eta$. This
induces the variation
\begin{equation}
\delta E_{\rm ex} = { 2 A_{\rm ex} \over m_{\rm s}^2 }
\int d^2 R \int_{z_{\rm L}}^{z_{\rm R}} dz
\, (\partial_z \bm\eta) \cdot (\partial_z {\bf M}).
\end{equation}
Integrating by parts gives
\begin{equation}
\delta E_{\rm ex}= { - 2 A_{\rm ex} \over m_{\rm s}^2 }
\int d^2 R
\left[
\int_{z_{\rm L}}^{z_{\rm R}} dz \,
\bm\eta \cdot \partial^2_z {\bf M}
- \left[
\bm\eta \cdot \partial_z {\bf M})
\right]\Big|_{z_{\rm L}}^{z_{\rm R}}
\right]
\end{equation}
or
\begin{eqnarray}
\delta E_{\rm ex} &=& { - 2 A_{\rm ex} \over m_{\rm s}^2 }
\int d^2 R
\int_{z_{\rm L}}^{z_{\rm R}} dz \,
\bm\eta \cdot
\left\{
\partial^2_z {\bf M}\right. \nonumber \\
&  & \left. \mbox{+ } \partial_z {\bf M} \left[\delta(z-{z_{\rm L}})
-  \ \delta(z-{z_{\rm R}})
\right]\right\}.
\end{eqnarray}
The exchange field
$H_{\rm ex}=-\gamma^{-1}(\delta E_{\rm ex}/ \delta {\bf M})$. Therefore, since
$\bm\eta\equiv \delta{\bf M}$,
\begin{equation}
{\bf H}_{\rm ex}
= { 2 A_{\rm ex} \over \gamma m_{\rm s}^2 } \left\{\partial^2_z {\bf M}
+ \partial_z{\bf M}
\left[
\delta(z-{z_{\rm L}}) \right. \right.
- \left. \left. \delta(z-{z_{\rm R}}) \right]
\right\}.
\label{zonly}
\end{equation}
The $\alpha=x$ and $\alpha=y$ components of (\ref{exenergy}) do not generate surface terms so
(\ref{zonly}) generalizes to
\begin{eqnarray}
{\bf H}_{\rm ex} &=&
{ 2 A_{\rm ex} \over \gamma m_{\rm s}^2 }
\nabla^2 {\bf M} \nonumber \\
&+&
{ 2 A_{\rm ex} \over \gamma m_{\rm s}^2 }
\left[\delta(z-z_{\rm L})-\delta(z-z_{\rm R})\right]\partial_z{\bf M}.
\end{eqnarray} \\

\appendix{\bf Appendix B: Transport Boundary Conditions} \\

The interfacial boundary conditions (\ref{BC}) for the drift-diffusion equations
(\ref{Fick}), (\ref{spincurrent}), (\ref{ferrocurrent}), and (\ref{ferrospincurrent})
can be derived most convincingly by averaging
the boundary conditions for the semi-classical distribution functions
$g_\uparrow({\bf k},{\bf r})$ and $g_\downarrow({\bf k},{\bf r})$ used in a Boltzmann equation
treatment of spin transport. \cite{Stiles:2003} A simpler alternative is to proceed
phenomenologically and write kinetic equations that connect the currents to the densities.
Temporarily, we assume that the ferromagnetic magnetization ${\bf M}=M{\bf u}$ is
parallel to the interface normal, i.e., parallel to the z-axis.

In linear approximation, the
longitudinal particle current that flows through the interface for each spin type is
\begin{eqnarray}
j_{\uparrow}&=&k_\uparrow({\mu}_\uparrow^{\bf NM}-{\mu}_\uparrow^{\bf FM})
=\textstyle{1\over2}k_\uparrow(a_0 \delta n_\uparrow^{\bf NM}-a_\uparrow\delta n_\uparrow^{\bf FM})  \nonumber \\ \label{aupdown} \\
j_{\downarrow}&=&k_\downarrow({\mu}_\downarrow^{\bf NM}-{\mu}_\downarrow^{\bf FM})
=\textstyle{1\over2}k_\downarrow(a_0 \delta n_\downarrow^{\bf NM}-a_\downarrow\delta n_\downarrow^{\bf FM}). \nonumber \\ \nonumber
\end{eqnarray}
Here, $k_\uparrow$ and $k_\downarrow$ are kinetic coefficients,
${\mu}_\uparrow^{\rm NM}$,$\,{\mu}_\downarrow^{\rm NM}$,$\,{\mu}_\uparrow^{\rm FM}$,
and ${\mu}_\downarrow^{\rm FM}$  are electrochemical potentials, and $a_0$, $a_\uparrow$,
and $a_\downarrow$ are {\it twice} the inverse density of states per spin at the Fermi
level for electrons in the non-magnet and
for majority and minority electrons in the ferromagnet, respectively.
Using (\ref{aupdown}),
the same algebra that led from (\ref{updowncurrent}) to (\ref{ferrocurrent})
now yields
\begin{eqnarray}
j&=&j_\uparrow+j_\downarrow \nonumber \\
& = & (k_\uparrow + k_\downarrow)a_0\delta n^{\rm NM}
-(a_\uparrow k_\uparrow + a_\downarrow k_\downarrow)\delta n^{\rm FM}  \nonumber \\
 & +& {2\over\hbar}(k_\uparrow - k_\downarrow)a_0 m_z -
{2\over \hbar}(a_\uparrow k_\uparrow - a_\downarrow k_\downarrow)\delta m.
\label{test}
\end{eqnarray}

The physical charge accumulation $\delta n_0$ at the interface
can be partitioned arbitrarily between $\delta n^{\rm NM}$ and $\delta n^{\rm FM}$.
If we choose $\delta  n =\delta n^{\rm NM}$ and $\delta n^{\rm FM}=0$,
the boundary condition for the particle current becomes
\begin{eqnarray}
\label{finalj}
j & =&  (k_\uparrow + k_\downarrow)a_0\delta n_0
+ 2\hbar^{-1}(k_\uparrow - k_\downarrow)a_0 m_z \nonumber \\
&-&2\hbar^{-1}(a_\uparrow k_\uparrow - a_\downarrow k_\downarrow)\delta m.  \\  \nonumber
\end{eqnarray}
In exactly the same way, we derive the boundary condition for the longitudinal spin current from
\begin{eqnarray}
Q_{zz}&=&j_\uparrow-j_\downarrow \nonumber \\
&=&  \textstyle{1\over2}\hbar(k_\uparrow - k_\downarrow)a_0\delta n
 + (k_\uparrow + k_\downarrow)a_0 m \nonumber \\
&-&
(a_\uparrow k_\uparrow + a_\downarrow k_\downarrow)\delta m.
\label{finalQQ}
\end{eqnarray}
Equations (\ref{finalj}) and (\ref{finalQQ}) reproduce the first
and last boundary conditions in (\ref{BC}) if we use the Boltzmann
equation treatment of the boundary conditions \cite{Stiles:2003}
to make the choice
\begin{equation}
k_\sigma ={A_{\rm FS} \over (2 \pi)^3 \hbar}{1-{\rm R}_\sigma \over 4 {\rm R}_\sigma}.
\label{kdef}
\end{equation}

\begin{figure}
\includegraphics{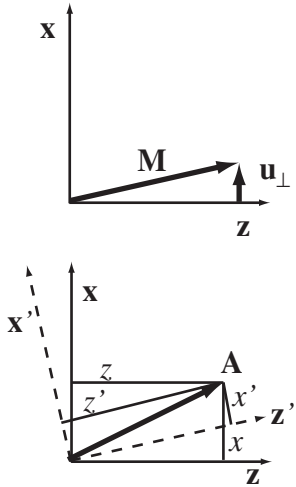}
\caption{The z-axis of unprimed coordinate system is perpendicular
to the material interface. The z-axis of the primed coordinate
system is parallel to the magnetization ${\bf M}$. }
\label{fig:rotation}
\end{figure}

We stated the boundary condition for the transverse spin current in Section III,
leaving a constant of proportionality $b$ unspecified in (\ref{nQBC}). As above,  we
choose this constant for consistency with the
Boltzmann equation treatment \cite{Stiles:2003}. The result is
\begin{equation}
{\bf Q}^{\rm NM}_{\perp z}={A_{\rm FS} \over (2 \pi)^3 \hbar}  a_0 {\bf m}_\perp,
\label{aperpBC}
\end{equation}
where $A_{\rm FS}$ is the area of the Fermi surface projected onto the interface.
The reflection coefficients in (\ref{kdef}) do not appear in (\ref{aperpBC}) despite the fact that
spins can rotate into the transverse direction upon quantum mechanical reflection
or transmission at an interface. This is so because the rotation effect disappears
after Fermi surface averaging. \cite{SZ:2002a}

Tserkovnyak and co-workers \cite{Tserkovnyak:2002} discuss another
contribution to $Q^{\rm NM}_{\alpha \beta}$ due to
``spin-pumping'' from an adjacent ferromagnet.  When the
magnetization is time-harmonic with frequency $\omega$, the time
derivative of the magnetization can have a component that is
transverse to the magnetization and there is an additional term in
(\ref{aperpBC})
\begin{equation}
{\bf Q}^{\rm NM}_{\perp z}={A_{\rm FS} \over (2 \pi)^3 \hbar}  a_0
{\bf m}_\perp + {A_{\rm FS} \over (2 \pi)^3 } {\hbar\omega\over 2}
{\bf u}_\perp , \label{spinPumping}
\end{equation}
For our choice of parameters, this new term is numerically much
smaller than the other term except for high frequencies and low
currents. Nevertheless, we have retained it for all the
calculations presented here.

As a final step, we must take explicit account of the fact that
${\bf M}$ is tilted slightly away from the
interface normal $\hat{\bf z}$. This is not difficult if we define a
a new coordinate system $x'y'z'$ where $\hat{\bf z}'$
points along ${\bf M}$ (Figure~\ref{fig:rotation}).
In that case, for example,
\begin{equation}
{\bf Q}^{\rm NM}_{\perp' z}={A_{\rm FS} \over (2 \pi)^3 \hbar}  a_0 {\bf m}_{\perp'},
\label{aperppBC}
\end{equation}
replaces (\ref{aperpBC}) as the boundary condition for the transverse spin current.
However, because the direction of ${\bf u}_\perp$ in Figure~\ref{fig:rotation} varies along the interface when
the excitation wave vector $K\neq 0$, it behooves us to evaluate (\ref{aperppBC})
using the original, interface-referenced coordinates $xyz$.

The transformation law
between the $xyz$ and $x'y'z'$ coordinate systems (Figure~\ref{fig:rotation})
simplifies considerably when  $u_\perp \ll 1$, as is the case for our problem. Then,
\begin{eqnarray}
A_x'&=&A_x-u_\perp A_z \nonumber \\
A_y'&=&A_y-u_\perp A_z \nonumber \\
A_z'&=&A_z,
\label{Rotate}
\end{eqnarray}
for any vector ${\bf A}$ that lies close to the $z$ axis,
This rule shows that (\ref{aperppBC}) is indeed identical to the transverse boundary
conditions stated in (\ref{BC}). It also shows that the longitudinal boundary
conditions stated above are unaffected by the fact that ${\bf u}_\perp \neq 0$. \\


\begin{thebibliography}{24}
\bibitem{Qz'z>0} The first term on the right side of (\ref{Qinmagnet}) shows that
$Q^{\rm FM}_{zz}$ is positive deep in the bulk where $\delta m=0$.
The diffusive correction can change its sign near the interface,
but this does not occur for the physical range of parameters we
use.
\end{thebibliography}
\end{document}